\documentclass[
  reprint,
  nofootinbib,
  amsmath,amssymb,
  aps,
  prd,
  floatfix,
  showkeys
]{revtex4-1}

\usepackage[USenglish]{babel}

\usepackage{amsmath}
\usepackage{amssymb}
\usepackage{amsfonts}
\usepackage{amsbsy}
\usepackage{bm}            
\usepackage{mathrsfs}
\usepackage{slashed}
\usepackage{extarrows}

\usepackage{graphicx}
\usepackage{graphics}
\usepackage{epstopdf}
\usepackage{dcolumn}       

\usepackage[dvipsnames]{xcolor}

\usepackage{array}
\usepackage{multirow}
\usepackage{enumitem}

\usepackage[caption=false]{subfig}

\usepackage{url}
\usepackage{varioref}
\usepackage{ulem}         
\usepackage{float}
\usepackage{verbatim}
\usepackage{latexsym}
\usepackage{orcidlink}
\usepackage{setspace}

\PassOptionsToPackage{
    colorlinks,
    citecolor=blue,
    urlcolor=magenta,
    linkcolor=blue
}{hyperref}
\usepackage{hyperref}

\newcommand{\GSSI}{Gran Sasso Science Institute (GSSI), I-67100 L’Aquila, Italy}
\newcommand{\GranSasso}{INFN, Laboratori Nazionali del Gran Sasso, I-67100 Assergi, Italy}

\newcounter{widefn}

\input epsf

\RequirePackage[colorlinks,citecolor=blue,urlcolor=magenta,linkcolor=blue]{hyperref}

\newcommand{\MBH}{M_{\text{BH}}}
\newcommand{\rout}{r_\text{o}}
\newcommand{\Lscale}{{\cal L}_\text{scale}}

\begin{document}
\tolerance=5000

\title{Axial tidal Love numbers of black holes in matter environments}

\author{Simone~D'Onofrio \orcidlink{0000-0001-7925-3905}}
\email{donofrio@ice.csic.es}
\affiliation{ Institute of Space Sciences (ICE, CSIC) C. Can Magrans s/n, 08193 Barcelona, Spain}

\author{Sayak Datta \orcidlink{0000-0002-4774-0298}}
\email{sayak.datta@gssi.it}
\affiliation{\GSSI}
\affiliation{\GranSasso}

\author{Andrea Maselli \orcidlink{0000-0001-8515-8525}}
\email{andrea.maselli@gssi.it}
\affiliation{\GSSI}
\affiliation{\GranSasso}

\date{\today}

\begin{abstract}
We study the axial (magnetic) tidal Love numbers 
of a Schwarzschild black hole surrounded by a 
spherically symmetric matter distribution. While 
the formalism developed here is general, we 
specialize to the case of anisotropic fluids as 
a proxy for dark matter distributions, computing 
the Love numbers for different density profiles 
of astrophysical interest. We employ two 
complementary methods: a small-compactness 
expansion, yielding closed-form analytic 
expressions, and direct numerical integration 
of the perturbation equations. 
We discuss the connection between different 
formulations of the fluid perturbations and the 
resulting Love numbers. We further show that density 
profiles lacking compact support generically 
produce logarithmic terms in the asymptotic 
expansion of the perturbation variable, which 
obstruct the standard tidal matching procedure 
and whose origin we trace to the absence of a 
strictly vacuum exterior. Our findings highlight 
the importance of controlling the asymptotic 
structure of the matter distribution when 
defining tidal observables for black holes 
dressed by matter, and provide a general 
framework that can be applied to other 
spherically symmetric environments.
\end{abstract}

\maketitle
\tolerance=5000

\section{Introduction}

The tidal response of compact objects to external gravitational fields 
is encoded in their {\it tidal Love numbers} (TLNs) \cite{Flanagan:2007ix}. 
In general relativity (GR), the relativistic theory of TLNs was first developed 
for nonspinning bodies \cite{Damour:2009vw,Binnington:2009bb,Hinderer:2007mb,
Hinderer:2009ca} and later extended to rotating configurations 
\cite{Pani:2015nua,Pani:2015hfa,Landry:2015cva,Damour:2012yf,Gagnon-Bischoff:2017tnz}.
For nonspinning objects, TLNs naturally separate according to the parity 
of the tidal perturbation. The {\it electric} type (even-parity or polar) TLNs 
describe induced mass multipole moments and admit a direct Newtonian 
analogue~\cite{Flanagan:2007ix}. In contrast, the {\it magnetic} type (odd-parity 
or axial) TLNs describe induced current multipole moments sourced by 
magnetic-type tidal fields, which have no Newtonian counterpart and are 
therefore a purely relativistic effect. TLNs enter as measurable parameters 
in the gravitational-wave signals emitted by coalescing binaries: in the 
post-Newtonian (PN) expansion of the waveform phase they appear at fifth 
PN order \cite{Flanagan:2007ix}, and their large magnitude makes them 
observationally accessible despite the formal smallness of the PN order \cite{LIGOScientific:2017vwq,LIGOScientific:2018hze,LIGOScientific:2020aai}. 
Magnetic TLNs contribute at 6PN order, with a subdominant role relative to 
the electric ones~\cite{Yagi:2013sva,Landry:2015cva,Pani:2015hfa}.

In GR, the TLNs of vacuum, asymptotically flat, nonrotating black holes (BHs) 
vanish identically~\cite{Binnington:2009bb,Hinderer:2009ca,
Goldberger:2009qd,DeLuca:2023mio,
Nair:2024mya,1983grr..proc.....D,Damour:2009va,Damour:2009vw,Gurlebeck:2015xpa,Porto:2016zng,LeTiec:2020spy,Chia:2020yla,LeTiec:2020bos,Charalambous:2021mea,Creci:2021rkz,Bonelli:2021uvf,Ivanov:2022hlo,Katagiri:2022vyz,Ivanov:2022qqt,Berens:2022ebl,Bhatt:2023zsy, Sharma:2024hlz}, a property traced to hidden ladder symmetries of the vacuum 
perturbation Eqs.~\cite{Hui:2020xxx,Hui:2021vcv,
Charalambous:2021kcz,Charalambous:2022rre,
Riva:2023rcm,Rai:2024lho,DeLuca:2025zqr,Ghosh:2026vig} 
(see \cite{Chakraborty:2026qru,Rodriguez:2026iot} for recent review on this topic). 
A measurement of nonzero TLNs in a compact binary would therefore signal a 
departure from the vacuum BH paradigm in GR, pointing to the possible existence 
of exotic compact objects \cite{Uchikata:2016qku,Cardoso:2017cfl,Maselli:2017vfi,Maselli:2017cmm,Cardoso:2019rvt,Katagiri:2024fpn,Berti:2024moe}, 
modifications of gravity \cite{Saffer:2021gak,Katagiri:2023umb,DeLuca:2023mio,
Garcia-Saenz:2025urd,Singha:2025xah,Cano:2025zyk}, 
or deviations of the surrounding spacetime from vacuum due to the presence of 
matter or astrophysical environments~\cite{Cardoso:2019upw,Cardoso:2021wlq,Katagiri:2023yzm,
Chakraborty:2024gcr,Cannizzaro:2024fpz,DeLuca:2022xlz,Chakravarti:2025awj}.

Astrophysical BHs are generically embedded in matter-rich environments, such as 
accretion disks  and dark matter halos, which can endow the system with nonzero 
TLNs. Understanding the impact of such environments on gravitational-wave 
signals has recently attracted considerable interest, particularly in view of next-generation 
detectors such as LISA \cite{2017arXiv170200786A}, TianQin \cite{TianQin:2015yph}, 
LGWA \cite{Ajith:2024mie} and the Einstein Telescope \cite{ET:2025xjr}, which will 
come online with dramatically improved sensitivity and promise to detect even tiny 
deviations from vacuum across a broad range of source masses \cite{Cardoso:2019rou}. 
Current observational data already place constraints on BH 
environments from stellar-mass binaries~\cite{Cardoso:2019upw,CanevaSantoro:2023aol, Andres-Carcasona:2025bni}. 
Due to the spatial extent and local density of surrounding matter, 
environmental TLNs can be large enough to significantly affect the emitted signals, 
potentially mimicking new-physics effects and introducing systematic biases in 
parameter estimation if not properly accounted for \cite{DeLuca:2025bph,Katagiri:2023yzm,Cannizzaro:2024fpz}.

Black holes embedded in matter environments are 
particularly relevant for binaries with large mass 
asymmetry, such as extreme and intermediate mass-ratio 
inspirals, where the massive primary is expected to 
retain its matter distribution throughout the inspiral, 
and measurements of its TLNs could allow tests 
deviations from vacuum 
GR~\cite{Pani:2019cyc,Datta:2021hvm,Piovano:2022ojl}.

A consistent theoretical framework for tidal deformability 
in nonvacuum settings is therefore essential both to 
characterize the properties of binary environments and 
to avoid contamination from environmental effects when 
searching for genuine new-physics signatures in 
strong-field gravitational-wave observations.

In this work, we study the axial TLNs of a nonrotating BH surrounded 
by a spherically symmetric anisotropic matter distribution. While the formalism is fully 
general, we specialize to density profiles motivated by dark matter halos, including 
the Hernquist, NFW, and Einasto models. We derive closed-form analytic 
expressions via a small halo-compactness expansion and validate them through 
direct numerical integration of the perturbation equations. These findings complement 
and extend previous studies of environmental TLNs~\cite{Cardoso:2019upw,Cardoso:2021wlq,
Chakraborty:2024gcr}.

We further clarify the ambiguities in the definition of axial TLNs for anisotropic fluid 
configurations, extending previous results for the isotropic case~\cite{Pani:2018inf}, 
and identify the physical origin of different prescriptions that have appeared in the 
recent literature. We also point out that density profiles lacking compact support 
generically produce logarithmic terms in the asymptotic expansion of the perturbation, 
which obstruct the standard matching procedure, and trace their origin to the absence 
of a strictly vacuum exterior region.

All analytical and numerical results presented in this work are publicly available in the 
\textit{GitHub} repository \cite{RepoGitHub,DattaGitHub,SGREP_REPO}.
Hereafter we use geometric units, such that $G=c=1$.

\section{Tidal response of anisotropic fluids}\label{sec:TidalResponse}

To investigate the impact of a spherically symmetric dark matter distribution on the 
BH geometry, we adopt the Einstein cluster prescription \cite{7bb06a79-8225-31c6-88c3-0c4f8a76b072,1970GReGr...1...19K, 1971GReGr...2..321B, 1968ApJ...153L.163Z,10.1098/rspa.1974.0065,Comer:1993rx, Magli:1997qf, Gair:2001qu, Szybka:2018hoe, Mahajan:2007vw}  as a concrete realization of a broader class of anisotropic fluid models. This framework has recently been employed as an effective model 
for dark matter halos surrounding both nonrotating and spinning BHs \cite{Lake:2006pp, Boehmer:2007az,Geralico:2012jt, Jusufi:2022jxu, Acharyya:2023rnq, Jusufi:2022jxu, Cardoso:2021wlq, Figueiredo:2023gas, Speeney:2024mas, Cardoso:2022whc, Datta:2023zmd, Shen:2023erj, Shen:2024qbb, Konoplya:2022hbl, Ovgun:2025bol,Fernandes:2025osu,Datta:2026krm}. 
In this framework, one considers a large ensemble of noninteracting particles 
moving on circular geodesics with isotropically distributed angular momenta. 
Upon coarse-graining, the collective dynamics of the ensemble can be 
described by an anisotropic fluid with vanishing radial pressure.

In this section, we briefly outline the formalism required to describe relativistic 
tidal perturbations in anisotropic fluids, focusing on axial perturbations. 
Some of the results obtained — in particular, the structure of the constraint 
equation for the fluid velocity perturbations and its physical implications — 
are independent of the specific matter model and the Einstein cluster 
prescription, and hold for any spherically symmetric anisotropic fluid.

\subsection{Background metric and axial perturbation}

We assume that the matter distribution is described 
by the stress-energy tensor
\begin{equation}
    T_{\mu\nu} = (\rho + p_t) u_\mu u_\nu + p_t g_{\mu\nu} + (p_r - p_t) k_\mu k_\nu \ ,
\end{equation}
where $u^\mu$ is the fluid four-velocity, normalized 
as $u^\mu u_\mu = -1$, and $k^\mu$ is a spacelike unit 
vector orthogonal to $u^\mu$, satisfying $k^\mu k_\mu = 1$ 
and $u^\mu k_\mu = 0$. The quantities $\rho$, $p_r$, and 
$p_t$ denote the energy density, radial pressure, and 
tangential pressure, respectively, and depend only on 
the radial coordinate.

We consider a static, spherically symmetric background 
BH spacetime with coordinates $x^\mu = (t,r,\theta,\phi)$ 
and line element
\begin{align}
    ds^2 = \bar{g}_{\mu\nu} dx^\mu dx^\nu 
    = -e^{\nu(r)} dt^2 + e^{\lambda(r)}dr^2 + r^2 d\Omega^2 
    \ .
\end{align}
In this background, the normalization and orthogonality conditions of $u^\mu$ and 
$k^\mu$ imply
\begin{equation}
    \bar{u}^\mu = \left(e^{-\nu/2}, 0, 0, 0 \right) \ , 
    \qquad 
    \bar{k}^\mu = \left(0, e^{-\lambda/2}, 0, 0 \right) \ .
\end{equation}
Here and in the following, an overbar denotes background 
quantities. We also introduce the mass function $m(r)$ 
defined through
\begin{equation}
    e^{-\lambda} = 1 - \frac{2m(r)}{r} \ .
\end{equation}

We now consider axial metric perturbations 
at linear order. In the Regge--Wheeler gauge 
\cite{Regge:1957td,Zerilli:1970se,Sago:2002fe}, the 
perturbed metric
\begin{equation}
g_{\mu\nu}(x^\mu) = \bar{g}_{\mu\nu}(x^\mu) + \delta g_{\mu\nu}(x^\mu) \ ,
\end{equation}
can be expanded in tensor harmonics as
\begin{equation}
    \delta g_{\mu\nu} = \sum_{\ell,m} 
    \left(\begin{matrix}
        0 & 0 & h_0^{\ell m}(t,r) S^{\ell m}_\theta & h_0^{\ell m}(t,r) S^{\ell m}_\phi\\
        * & 0 & h_1^{\ell m}(t,r) S^{\ell m}_\theta & h_1^{\ell m}(t,r) S^{\ell m}_\phi \\
        * & * & 0 & 0 \\
        * & * & * & 0 \\ 
    \end{matrix}\right) \ ,
\end{equation}
where $\ell = 2,\ldots,\infty$, $-\ell \le m \le \ell$ 
denotes the azimuthal index (not to be confused with 
the mass function $m(r)$), and a star $*$ denotes the 
entry related by symmetry of the metric perturbation. 
The odd-parity vector spherical harmonics are defined as
\begin{equation}
    (S_\theta^{\ell m}, S_\phi^{\ell m}) 
    = \left(-\frac{1}{\sin\theta}\partial_\phi Y^{\ell m}, 
    \sin\theta\,\partial_\theta Y^{\ell m} \right) \ ,
\end{equation}
with $Y^{\ell m}(\theta,\phi)$ the scalar spherical harmonics. 

In the axial sector, metric perturbations do not couple 
to pressure and density perturbations, but they do 
couple to fluid velocities. For the four-velocity perturbation 
and the spacelike vector $k^\mu$, we introduce the 
functions $U_{\ell m}(r,t)$ and $U_{k,\ell m}(r,t)$ 
such that
\begin{align}
\delta u^t &= \delta u^r = 0 \ ,\\
\delta u^\theta &= -\frac{e^{\nu/2}}{4 \pi (\bar{\rho} + \bar{p}_t) r^2\sin\theta} \sum_{\ell,m}U_{\ell m}(t,r) Y_{\ell m,\phi}\ ,\\
\delta u^\phi &= \frac{e^{\nu/2}}{4 \pi (\bar{\rho} + \bar{p}_t) r^2 \sin\theta} \sum_{\ell,m}U_{\ell m}(t,r) Y_{\ell m,\theta}\ ,
\end{align}
and
\begin{align}
\delta k^t &= \delta k^r = 0 \ ,\\
\delta k^\theta &= -\frac{e^{-\lambda/2}}{4 \pi (\bar{\rho} + \bar{p}_t) r^2\sin \theta} \sum_{\ell,m} U_{k,\ell m}(t,r) Y_{\ell m,\phi} \ ,\\
\delta k^\phi &= \frac{e^{-\lambda/2}}{4 \pi (\bar{\rho} + \bar{p}_t) r^2 \sin\theta}\sum_{\ell,m}U_{k,\ell m}(t,r) Y_{\ell m,\theta} \ .
\end{align}
Owing to spherical symmetry, the perturbation equations 
are independent of the azimuthal index $m$. 
For notational simplicity, hereafter we will 
also suppress the multipolar index $\ell$.

The Einstein equations sourced by the anisotropic fluid 
are expanded to first order as
\begin{equation}
G_{\mu\nu}[\bar{g}_{\alpha\beta}] + \delta G_{\mu\nu}[\delta g_{\alpha\beta}] 
= \bar{T}_{\mu\nu} + \delta T_{\mu\nu} \ ,
\label{eq:expfieldeqs}
\end{equation}
where $\delta G_{\mu\nu}$ denotes the linearized Einstein 
operator acting on $\delta g_{\mu\nu}$. At the background 
level, the field equations reduce to
\begin{align}
    \nu' &= \frac{2m + 8\pi \bar{p}_r r^3}{r(r - 2m)} \ , \label{TOV_nu'}\\
    m' &= 4\pi r^2 \bar{\rho} \ , \label{TOV_m'} \\
    \bar{p}_r' &= -\frac{(\bar{\rho} + \bar{p}_r)\left(m + 4\pi r^3 \bar{p}_r\right)}{r(r - 2m)} 
    + \frac{2}{r}(\bar{p}_t - \bar{p}_r) \ , \label{TOV_pr'}
\end{align}
where a prime denotes differentiation with respect to $r$.

\subsection{Equations governing the tide}

The axial sector of the gravitational perturbations 
is governed by the following system of second-order 
partial differential equations:
\begin{widetext}
\begin{align}
    &e^{-\nu}\dot{h}_0 - e^{-\lambda} h_1' - \frac{1}{r^2}\Big(2m - 4\pi r^3 (\bar{\rho} - \bar{p}_r)\Big)h_1 = 0 \label{masterEq_U_Uk1}\\ 
    &e^{-\nu} \Big(\dot{h}_0' - \Ddot{h}_1\Big) - \frac{2e^{-\nu}}{r} \dot{h}_0-\left(\frac{(\ell-1)(\ell+2)}{r^2} + 16\pi (\bar{p}_t-\bar{p}_r)\right) h_1 = - \frac{4 (\bar{p}_t - \bar{p}_r)}{\bar{\rho} + \bar{p}_t} U_k \label{masterEq_U_Uk2}\\
    & e^{-\lambda} \Big(h_0'' - \dot{h}'_1\Big) - 4 \pi r (\bar{\rho} + \bar{p}_r)\Big(h_0' - \dot{h}_1\Big) - \frac{2}{r}e^{-\lambda}\dot{h}_1 - \frac{1}{r^2}\left(\ell (\ell +1) - \frac{4m}{r} + 8\pi r^2 (\bar{\rho} + 2 \bar{p}_t - \bar{p}_r)\right)h_0 = -4 e^{\nu} U \label{masterEq_U_Uk3}\ ,
\end{align}
\end{widetext}
Eqs.~\eqref{masterEq_U_Uk1}--\eqref{masterEq_U_Uk3} follow from the 
$\theta\theta$, $r\theta$, and $(t\theta,t\phi)$ components of the linearized 
Einstein equations, $\delta G_{\mu\nu} = \delta T_{\mu\nu}$, respectively. 
A dot denotes partial differentiation with respect to $t$. In the isotropic limit 
$\bar{p}_t = \bar{p}_r$, Eqs.~\eqref{masterEq_U_Uk1}--\eqref{masterEq_U_Uk3} 
reduce to the axial perturbation equations for a perfect fluid exhibited in 
Ref.~\cite{Pani:2018inf}.

The system involves four unknown functions, $h_0$, $h_1$, $U$, and $U_k$. 
From the $\theta$ component of the linearized conservation equation 
$\nabla_\mu T^{\mu\nu} = 0$ we obtain a relation between the velocity 
perturbations\footnote{Equivalently, this relation can also be found by 
rearranging the system of Eqs.~\eqref{masterEq_U_Uk1}--\eqref{masterEq_U_Uk3}.}:
\begin{align}\label{U-Uk_relation}
    \partial_t \Big(U &- 4\pi e^{-\nu}(\bar{\rho}+\bar{p}_t) h_0 \Big)\nonumber\\
    &+\frac{\bar{p}_r-\bar{p}_t}{e^\lambda(\bar{\rho}+\bar{p}_t)}\Big[
\partial_r\Big(U_k - 4 \pi (\bar{p}_t + \bar{\rho})h_1\Big) \\
&+ \mathcal{F}(r)\Big(U_k - 4 \pi(\bar{p}_t + \bar{\rho})h_1\Big) \Big]\nonumber= 0 \ ,
\end{align}
with 
\begin{align}
    \mathcal{F}(r)=&\frac{m (-\bar{p}_r+2 \bar{p}_t+\bar{\rho})+4 \pi  r^3 (\bar{p}_r (\bar{p}_t+2 \bar{\rho})-\bar{p}_t \bar{\rho})}{r (2 m-r) (\bar{p}_r-\bar{p}_t)}\nonumber\\
    &-\frac{(\bar{p}_r+\bar{\rho}) \bar{p}_t'}{(\bar{p}_r-\bar{p}_t) (\bar{p}_t+\bar{\rho})}-\frac{\bar{\rho}'}{\bar{p}_t+\bar{\rho}}\ .
\end{align}
An additional condition is required to close the system. Here, we assume 
that the fluid is irrotational, i.e.\ that the vorticity of $u^{\mu}$
\begin{equation}\label{vorticity}
    \omega^\alpha[u^{\gamma}] = \frac{1}{2}\epsilon^{\alpha \beta \mu\nu}u_{\beta;\mu}u_\nu \ ,
\end{equation}
vanishes. The condition $\omega^\alpha[u^{\gamma}]=0$ is equivalent to
\begin{equation}\label{U_irrotational}
    U = 4\pi e^{-\nu}(\bar{\rho}+\bar{p}_t) h_0 \ .
\end{equation}

Substituting this relation into Eq.~\eqref{U-Uk_relation} yields a constraint 
equation for $U_k$,
\begin{equation}
\partial_r\Big(U_k - 4 \pi (\bar{p}_t + \bar{\rho})h_1\Big)
+ \mathcal{F}(r)\Big(U_k - 4 \pi(\bar{p}_t + \bar{\rho})h_1\Big)=0 \ .\label{eq:constrUk}
\end{equation}

Imposing that $u^\mu$ be vorticity-free does not uniquely determine $U_k$, 
but instead constrains the combination $U_k - 4 \pi (\bar{p}_t + \bar{\rho})h_1$. 
In the following, we focus on the particular branch
\begin{equation}\label{Uk_irrotational}
    U_k = 4\pi (\bar{\rho}+\bar{p}_t) h_1 \ ,
\end{equation}
which identically satisfies Eq.~\eqref{eq:constrUk}.

Interestingly, this choice also implies $\omega^{\alpha}[k^{\gamma}]=0$. 
Thus, while the irrotationality of $u^\mu$ does not by itself enforce the 
irrotationality of $k^\mu$, it admits a consistent branch in which both 
vector fields are vorticity-free.
Alternative branches are also possible. In particular, one may consider 
configurations with $U_k=0$ while still imposing Eq.~\eqref{U_irrotational}. 
In this case, the constraint Eq.~\eqref{eq:constrUk} is satisfied once 
the Einstein equations are taken into account, corresponding to a different 
physical realization of the fluid perturbations.

Substituting the expressions for $U$ and $U_k$ into 
Eqs.~\eqref{masterEq_U_Uk1}--\eqref{masterEq_U_Uk3}, we obtain
\begin{align}
    &e^{-\nu}\dot{h}_0 - e^{-\lambda} h_1' - \frac{1}{r^2}\Big(2m - 4\pi r^3 (\bar{\rho} - \bar{p}_r)\Big)h_1 = 0 \label{masterEq_anis1}\\ 
    &e^{-\nu} \Big(\dot{h}_0' - \Ddot{h}_1\Big) -  \frac{2e^{-\nu}}{r} \dot{h}_0-\frac{(\ell-1)(\ell+2)}{r^2} h_1 = 0 \label{masterEq_anis2}\\
    & e^{-\lambda} \Big(h_0'' - \dot{h}'_1\Big) - 4 \pi r (\bar{\rho} + \bar{p}_r)\Big(h_0' - \dot{h}_1\Big) - \frac{2}{r}e^{-\lambda}\dot{h}_1 \nonumber\\
    & \quad - \frac{1}{r^2}\left(\ell (\ell +1) - \frac{4m}{r} - 8\pi r^2 (\bar{\rho} + \bar{p}_r)\right)h_0 = 0 \label{masterEq_anis3}\ ,
\end{align}
For this branch, the vorticity-free condition removes the explicit dependence 
on the tangential pressure from the perturbation equations, leaving only the 
energy density and radial pressure, both directly and through the mass function 
via Eq.~\eqref{TOV_m'}.

Restricting to the stationary sector, Eq.~\eqref{masterEq_anis3} provides a 
second-order differential equation for $h_0$:
\begin{equation}\label{h0_master_irr_anis}
\begin{aligned}
    &e^{-\lambda} h_0'' - 4 \pi r (\bar{\rho} + \bar{p}_r)h_0' \\
    &- \frac{1}{r^2}\left(\ell (\ell +1) - \frac{4m}{r} - 8\pi r^2 (\bar{\rho} + \bar{p}_r)\right)h_0 = 0 \ ,
\end{aligned}
\end{equation}
which agrees in the isotropic limit with the equation derived in 
Refs.~\cite{Damour:2009vw,Pani:2018inf}.

As a final remark, the above irrotational configuration does not reduce to the 
strictly static case, which would instead require $U = U_k = 0$ together with 
vanishing time derivatives of all perturbation variables. This parallels the 
result of Ref.~\cite{Pani:2018inf} for isotropic fluids, where the magnetic TLNs 
obtained in the irrotational setup differ from those computed in the strictly 
static configuration.

The present analysis further clarifies the interplay between the velocity 
perturbations. In particular, imposing the irrotational condition on $k^\mu$, 
i.e.\ fixing $U_k$ as in Eq.~\eqref{Uk_irrotational}, constrains the evolution 
of $U$ through Eq.~\eqref{U-Uk_relation}. Consequently, simultaneously setting 
$U=0$ is incompatible with a dynamical configuration, and can only be 
consistently realized in the strictly static limit. 

We remark that the above results are general and do not depend on the specific 
form of the matter distribution, but only on the assumption of spherical symmetry 
and on the anisotropic fluid description.

\subsection{Regge--Wheeler equation}

Although the analysis presented in this work does not rely on the Regge--Wheeler formulation, it is useful to recast the perturbation equations in this framework to facilitate comparison with previous results in the literature. 

To this end, we rewrite 
the system of Eqs.~\eqref{masterEq_U_Uk1}--\eqref{masterEq_U_Uk3} 
as a single equation in terms of the Regge--Wheeler 
function $\psi(r,t)$, defined as \cite{Ferrari:2020nzo}
\begin{equation}\label{RW_psi}
    \psi = \frac{ e^{-(\lambda-\nu)/2}}{r} h_1 \ . 
\end{equation}
In this formulation, Eq.~\eqref{masterEq_anis1} 
becomes
\begin{equation}\label{RW_h0_psi}
    \Dot{h}_0  = e^{(\nu-\lambda)/2}\big(r \, \psi)' \ .
\end{equation}
Introducing the Fourier transform
\begin{equation}
h_i(t,r) = \int d\omega \, h_i(\omega,r) e^{-i\omega t}\ ,
\end{equation}
and substituting Eqs.~\eqref{RW_psi}--\eqref{RW_h0_psi} into 
Eq.~\eqref{masterEq_U_Uk2}, we obtain
\begin{equation}\label{RW_anisotropic}
    \frac{d^2\psi}{dr^{2}_\star} + \Big(\omega^2 - V_\text{eff}\Big)\psi 
    =  - 4e^\nu\frac{e^{(\nu-\lambda)/2}(\bar{p}_t - \bar{p}_r)}{r(\bar{\rho} + \bar{p}_t)} U_k  \ ,
\end{equation}
where the tortoise coordinate is defined by 
$dr/dr_\star = e^{(\nu-\lambda)/2}$, and the 
effective axial potential reads
\begin{equation}
    V_\text{eff} = e^\nu \left(\frac{\ell(\ell+1)}{r^2} - \frac{6m}{r^3} + 4\pi (\bar{\rho}+4\bar{p}_t - 5\bar{p}_r)\right) \ .
\end{equation}
The Regge--Wheeler equation in the irrotational case 
is obtained by substituting Eq.~\eqref{Uk_irrotational} 
and using the definition \eqref{RW_psi}. In this case,
\begin{equation}\label{RW_irr}
    \frac{d^2\psi}{dr^{2}_\star} + \Big(\omega^2 - V^\text{irr}_\text{eff}\Big)\psi = 0 \ ,
\end{equation}
with
\begin{equation}
    V_\text{eff}^\text{irr} = e^\nu \left(\frac{\ell(\ell+1)}{r^2} - \frac{6m}{r^3} + 4\pi (\bar{\rho} - \bar{p}_r)\right) \ .
\end{equation}

Eq.~\eqref{RW_irr} agrees with the results of 
Ref.~\cite{Damour:2009vw} in the $\omega \to 0$ limit, 
and with those of Ref.~\cite{Cardoso:2021wlq} in the 
case of vanishing radial pressure $\bar{p}_r = 0$. 

This reformulation also helps to connect with some previous 
results in the literature. In Ref.~\cite{Chakraborty:2024gcr}, 
axial tidal perturbations of a nonspinning BH surrounded 
by a Hernquist dark matter profile (see Sec.~\ref{sec:HernAnalytic}) 
were recently computed, yielding two families of Love 
numbers, labelled \textit{``down''} and \textit{``up''}. 
Within our framework, this distinction can be traced 
back to the nonuniqueness of the solution of the 
constraint Eq \eqref{U-Uk_relation}, whose different 
branches correspond to distinct physical realizations 
of the fluid perturbations. In particular, we find that 
the \textit{``down''} prescription is physically equivalent 
to the irrotational branch considered here, while the 
strictly static limit — although not analyzed in detail 
in this work — corresponds to the \textit{``up''} Love 
numbers of \cite{Chakraborty:2024gcr} 
(see Eqs.~(45)-(46) in \cite{Chakraborty:2024gcr}). 

\subsection{Tidal Love Numbers}\label{sec:TLNDefinition}

We now consider the tidal response of the black hole–matter system. In the presence of an external tidal field, current multipole moments are induced. In the linear-response regime, the Love numbers are defined as the constants of proportionality 
relating the induced moments to the applied tidal field 
\cite{Binnington:2009bb,Damour:2009vw,Hinderer:2007mb}. For axial 
perturbations, we write
\begin{equation}
    S_{L} = \sigma_\ell H_{L} \ ,
    \label{tidal_sigma_def}
\end{equation}
where $\sigma_\ell$ is the magnetic deformability of multipolar order $\ell\ge 2$, 
while $S_{L}$ and $H_{L}$ denote the induced current multipole moment and the 
external magnetic tidal moment, 
respectively.\footnote{Here uppercase Latin indices are used as a shorthand 
for multi-indices $a_1\ldots a_\ell$.}

In geometric units, the tidal deformability has dimensions 
$[\sigma_\ell] = [\mathrm{length}]^{2\ell+1}$. It is then convenient to 
introduce the dimensionless TLNs
\begin{equation}
    \Tilde{k}_\ell^B = \frac{4(\ell+2)(2\ell-1)!!}{\ell-1}\frac{\sigma_\ell}{\Lscale^{2\ell+1}} \ ,
    \label{Axial_TLN_bar_def}
\end{equation}
where $\Lscale$ denotes a 
characteristic scale of the system. 
In the following, we present our results in terms 
of $\tilde{k}_\ell^B$, taking $\mathcal{L}_\text{scale}=M_\text{tot}$, 
namely the total mass of the BH plus its surrounding 
matter distribution \cite{Cardoso:2017cfl}. For $\ell=2$, 
this corresponds to $\tilde{k}_2^B = 48\,\sigma_2/M_\text{tot}^5$. 
This choice allows a direct comparison with previous 
studies of environmental TLNs~\cite{Cardoso:2017cfl,Chakraborty:2024gcr}.

The multipole moments introduced in Eq.~\eqref{tidal_sigma_def} can be read 
off from the asymptotic expansion of the spacetime metric at spatial infinity. 
We follow the prescription of Thorne \cite{Thorne:1980ru}, whereby the metric 
is written in an asymptotically Cartesian and mass-centered (ACMC) frame. In 
the axial sector, the relevant metric component takes the form
\begin{equation}
    g_{t\phi} = \sum_{\ell\geq2}\left(\frac{1}{r^{\ell}}\frac{4(2\ell-1)!!}{(\ell+1)!}S_{\ell m}
    - r^{\ell+1}\frac{1}{(\ell+1)!}H_{\ell m}\right)S^{\ell m}_{\phi} \ ,
    \label{ACMC_tphi}
\end{equation}
where the induced and tidal moments have been decomposed in terms of symmetric 
trace-free tensors as
\begin{equation}
    S_L = \sum_m S_{\ell m}\mathcal{Y}^{\ell m}_L \ , \qquad
    H_L = \sum_m H_{\ell m}\mathcal{Y}^{\ell m}_L \ ,
\end{equation}
with $Y_{\ell m} = \mathcal{Y}^{\ell m}_L n^L$, $r^2=\delta_{ij}x^i x^j$, $n^L=n^{i_1}n^{i_2}...n^{i_\ell}$
and $n^i=x^i/r$ \cite{Abdelsalhin:2019ryu,1967ApJ...149..591T}. 
Eq.~\eqref{ACMC_tphi} clearly separates the decaying and growing radial 
behaviors of the metric perturbation, associated with the induced multipole 
moments of the central object and the external tidal moments, respectively.

To assess how the axial Love numbers depend on the 
properties of the surrounding matter, we adopt density 
profiles motivated by models of dark matter halos at 
galactic centers~\cite{1969Afz.....5..137E,
1965TrAlm...5...87E,Navarro:1996gj,1990ApJ...356..359H}. 
In these environments, adiabatic accretion onto 
massive black holes is predicted to reshape the surrounding 
matter distribution, forming overdense spikes whose density 
scales with the BH mass and peaks close to the central 
object~\cite{Gondolo:1999ef,Sadeghian:2013laa}. Such 
configurations are phenomenologically most relevant for 
binaries with large mass asymmetry, such as extreme and 
intermediate mass-ratio inspirals, where the lighter 
companion acts as a weak perturbation that does not 
significantly disrupt the matter distribution~\cite{Kavanagh:2020cfn}, 
and where the tidal deformability of the central object offers 
unique potential to probe deviations from 
vacuum~\cite{Pani:2019cyc,Datta:2021hvm, Piovano:2022ojl}. By contrast, 
comparable-mass mergers are expected to deplete the 
matter environment through tidal forces and the merger 
process itself~\cite{DeLuca:2022xlz}, so that such systems 
may effectively appear as naked black holes by the time they 
enter the gravitational-wave observation band.

Within this framework, some care is required: realistic 
dark matter halos are expected to contain particles on 
orbits spanning a broad range of eccentricities~\cite{Vicente:2025gsg}, 
and therefore cannot be strictly described within the Einstein 
cluster prescription. Nevertheless, we expect the present 
analysis to provide qualitative insight into the tidal 
response induced by extended matter distributions.

\begin{figure}[H]
\begin{center}
\centering
\includegraphics[width=1\linewidth]{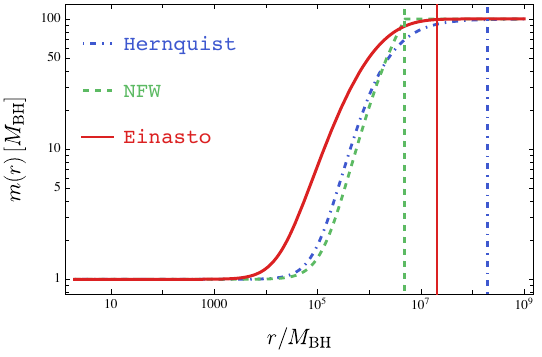}
\caption{Mass function $m(r)$ for the Hernquist, NFW, and Einasto profiles, together with the 
corresponding values of $R_{99}$, indicated by the vertical lines. We set $\MBH=1$, $M=100\MBH$, 
and $a_0=10^6\MBH$; for the NFW profile we also fix $r_c=5a_0$.}
\label{fig:mass-radius_profiles}
\end{center}
\end{figure}
%

\subsection{Matter profiles and numerical setup}\label{sec:DM_profiles}

We consider a family of profiles described by the parametric form 
\cite{Salucci:2018hqu}
\begin{equation}\label{parametric_density}
    \rho_\text{DM}(r) = \rho_0 \left(\frac{r}{a_0}\right)^{-\gamma}
    \left[1+\left(\frac{r}{a_0}\right)^\alpha\right]^{(\gamma-\beta)/\alpha} \ ,
\end{equation}
where the triplet $(\alpha,\beta,\gamma)$ selects a specific profile. The parameter $a_0$ sets the characteristic length scale at which $\rho_\text{DM}(a_0) = 2^{(\gamma-\beta)/\alpha}\rho_0$. The Hernquist and Navarro--Frenk--White (NFW) profiles correspond to $(\alpha,\beta,\gamma) = (1,4,1)$ and $(1,3,1)$, respectively. In addition, we consider the Einasto profile, which has a different functional form,
\begin{equation}\label{rho_DM_Einasto}
    \rho_\text{DM}(r) = \rho_0 
    \exp\left\{-d\left[\left(\frac{r}{a_0}\right)^{1/n} - 1\right]\right\} \ ,
\end{equation}
where $a_0$ plays the role of the effective radius in the original parametrization and we fix $n=6$, $d=53/3$ \cite{Graham:2005xx,Prada:2005mx}. For all profiles, we work in the physically viable regime 
$M_{\rm BH}\ll M\ll a_0$, in which the horizon of the BH coincides with the Schwarzschild one \cite{Cardoso:2021wlq,Figueiredo:2023gas}.

Both Newtonian and relativistic analyses predict that the dark matter density vanishes at the BH horizon. To enforce this behavior, we rescale the 
density as\footnote{A more realistic prescription would require that the density vanish at $r=4M_{\rm BH}$, as found in the relativistic analysis of Ref.~\cite{Sadeghian:2013laa}. Our approach can be 
straightforwardly generalized to this case following the procedure outlined below.}
\begin{equation}\label{rho_tot_w_DM}
    \bar{\rho}(r) \rightarrow \left(1 - \frac{2M_{\rm BH}}{r}\right)\rho_\text{DM}(r) \ .
\end{equation}
Within the Einstein cluster prescription we set $\bar{p}_r=0$ \cite{Cardoso:2019upw,Cardoso:2021wlq}. Eq.~\eqref{TOV_pr'} then determines the background tangential pressure in terms of the energy density,
\begin{equation}
    \bar{p}_t = \frac{m}{2(r-2m)}\,\bar{\rho} \ .
\end{equation}
In the case of vanishing background radial pressure, the master 
equation~\eqref{h0_master_irr_anis} becomes
\begin{equation}\label{h0_master_Ecluster}
   \left(1-\frac{2m}{r}\right) h_0'' - \frac{m'}{r}h_0' - \left(\frac{\ell(\ell+1)}{r^2} - \frac{4m}{r^3}-\frac{2m'}{r^2}\right) h_0 = 0 \ .
\end{equation}
We solve Eq.~\eqref{h0_master_Ecluster} using two complementary approaches. First, we perform a small-compactness expansion, which yields closed-form 
analytic expressions for the Love numbers. Second, we integrate the full master equation numerically. In both cases, the Love numbers are extracted by matching the solution at large radii to the ACMC asymptotic expansion~\eqref{ACMC_tphi} \cite{Hinderer:2007mb}. 
Although the formalism and the numerical pipeline developed here are fully general, in the following we focus on the dominant $\ell = 2$ contribution to the axial Love 
numbers.

\section{Results}

\subsection{Small compactness expansion}\label{Sec:smallCexpansion}

To obtain an analytic expression for $\tilde{k}_2^B$, we 
follow Refs.~\cite{Cardoso:2022whc,Cardoso:2019upw}. We  expand both the metric perturbation and the master Eq.~\eqref{h0_master_Ecluster} to linear order in the compactness ${\cal C} = M/a_0$, e.g.\ $h_0 = h_0^{(0)} + {\cal C}\, h_0^{(1)}$, with $h_0^{(0)}$ corresponding to the vacuum solution. The latter satisfies
\begin{equation}\label{eqh0_zeroOrder}
    h_0^{(0)}{}'' - \frac{1}{r^2}\left(6 + \frac{8\MBH}{r-2\MBH}\right)h_0^{(0)} = 0 \ .
\end{equation}
At zeroth order in the compactness, matching to 
the asymptotic form of the metric~\eqref{ACMC_tphi} 
and imposing regularity at the BH horizon fixes 
the coefficient of the decaying mode to zero, consistent with the vanishing of the Love number of an isolated Schwarzschild BH. The resulting solution is
\begin{equation}\label{h0_BH}
    h_0^{(0)} = -\frac{H_{20}}{6}r^3\left(1-\frac{2\MBH}{r}\right) \ .
\end{equation}
Eq.~\eqref{h0_master_Ecluster}, and hence the first-order correction $h_0^{(1)}$, depends only on the mass function $m(r)$ and its radial derivative. For each profile, the mass function is obtained by solving the background field Eq,~\eqref{TOV_m'}, supplemented by the condition $m(2M_{\rm BH}) = M_{\rm BH}$ and by the asymptotic condition at large radii, $m(r\rightarrow\infty)=M+\MBH$.

Figure~\ref{fig:mass-radius_profiles} shows the mass function for the three profiles considered here, for a representative choice of matter parameters.

\subsubsection{Hernquist profile}\label{sec:HernAnalytic}

The mass function of the Hernquist profile takes a particularly simple analytic form:
\begin{equation}\label{m_Hernquist}
    m(r) = \MBH + M \left(\frac{r-2\MBH}{r+a_0}\right)^2 \ .
\end{equation}
The first-order axial perturbation obeys the inhomogeneous equation
\begin{align}\label{h0_master_eta_first}
    h_0^{(1)}{}'' &- \frac{1}{r^2}\left(6 + \frac{8\MBH}{r-2\MBH}\right)h_0^{(1)}  = \mathcal{S} \ ,
\end{align}
where the source term is
\begin{equation}\label{S_Hernquist}
   \mathcal{S}= -\frac{ H_{20}}{3}\frac{a_0 r (4r^2 - 6 \MBH r + 5 a_0 r - 8 a_0 \MBH)}{(r+a_0)^3} \ .
\end{equation}
Eq.~\eqref{h0_master_eta_first} can be solved 
using a Green's function approach. The associated 
homogeneous problem admits two linearly independent 
solutions:
\begin{align}
    \Psi_- =& A_1 r^3\left(1-\frac{2\MBH}{r}\right) \label{auto_functions-}\ ,\\ 
    \Psi_+ =& -\frac{A_2}{24 \MBH^5 r} \nonumber\\
    &\times\Bigg[2\MBH\left(3r^3 - 3\MBH r^2 - 2\MBH^2 r - 2\MBH^3\right) \nonumber\\
    &+ 3 r^4 \left(1-\frac{2\MBH}{r}\right)\log\left(\frac{r}{r-2\MBH}\right)\Bigg] \ ,\label{auto_functions+}
\end{align}
where $A_{1,2}$ are integration constants. The full solution is
\begin{align}
    h_0^{(1)}(r) =& \ \Psi_+(r)\int^{r}_{2\MBH} dr' \frac{\mathcal{S}(r') \Psi_-(r')}{W(r')} \nonumber\\
    &+ \Psi_-(r)\int_r^\infty dr'\frac{\mathcal{S}(r') \Psi_+(r')}{W(r')} \ ,\label{h0_Wronskian}
\end{align}
where the Wronskian is constant,
\begin{equation}
    W(r) = \Psi_+'\Psi_- - \Psi_+\Psi_-' = A_1 A_2 \ .
\end{equation}
At large radius, the perturbation~\eqref{h0_Wronskian} 
admits a series expansion of the form
\begin{equation}\label{h0_expansion_infinity}
    h_0(r) \sim \sum_{i=-\infty}^3 a_i r^i + \log \left(\frac{r}{R_s}\right)\sum_{i=2}^{\infty} b_i r^{-i} \ ,
\end{equation}
where $R_s$ is an arbitrary length scale introduced to render the argument 
of the logarithm dimensionless, and the coefficients $(a_i,b_i)$ are fully 
determined by the integrals in Eq.~\eqref{h0_Wronskian}. The logarithmic 
terms arise from the first integral in Eq.~\eqref{h0_Wronskian}. While such 
terms were previously identified in a specific setup in~\cite{Cardoso:2019upw}, 
their physical interpretation remained unclear. We interpret this behavior as 
a consequence of the asymptotic structure of the Hernquist profile, which 
lacks compact support and never becomes strictly vacuum at large radii, 
producing logarithmic contributions upon integration. As a result, the expansion 
in Eq.~\eqref{h0_expansion_infinity} does not satisfy the requirements of the 
ACMC matching. For the moment, and to facilitate 
comparison with previous work~\cite{Chakraborty:2024gcr,Cardoso:2021wlq}, 
we proceed by neglecting the logarithmic terms. 
Note however that this generically 
leads the Love number to depend on the arbitrary scale $R_s$, 
a dependence that we believe is unphysical and is resolved by 
the cutoff prescription of Sec.~\ref{subsec:analytic_Hernquist_cutoff} 
which eliminates the logarithmic terms, and restores 
agreement with the numerical results.

Matching the expansion in Eq.~\eqref{h0_expansion_infinity} 
to Eq.~\eqref{ACMC_tphi} to identify the growing and decaying 
modes, we obtain the axial Love number for the Hernquist profile 
at first order in ${\cal C}$:
\begin{align}\label{k2_analytical}
    \Tilde{k}^B_2=&\frac{4a_0^2 M}{25 \Lscale^5}\Big[25 a_0^2 + 114 a_0 \MBH + 108 \MBH^2 \nonumber\\
    &\textstyle+20(a_0 + 2\MBH) (5a_0 + 6 \MBH) \log \left(\frac{R_s}{a_0 + 2 \MBH}\right)\Big] \ .
\end{align}

As expected, $\tilde{k}^B_2$ vanishes in the limit 
$M\rightarrow0$, recovering the result for an isolated 
Schwarzschild. In the limit $a_0, R_s\gg\MBH$, this reduces 
to
\begin{equation}
    \Tilde{k}_2^B \sim \frac{4\, a_0^4 M \left(1 - 4\log(a_0/R_s)\right)}{\Lscale^5} \ .
\end{equation}
Choosing $R_s=\Lscale$, our result agrees with 
\cite{Chakraborty:2024gcr} (see Eq.~(130) therein\footnote{There is a difference in 
the overall normalization due to the definition of the Love numbers. Here 
we follow the prescription of \cite{Damour:2009vw}. Accounting for the different 
definition of the multipole moments, the Love numbers are related by 
$k_\ell^\text{our} = 
\frac{2(\ell+1)(\ell+2)}{\ell(\ell-1)} k_\ell^\text{\cite{Chakraborty:2024gcr}}$.}).

\subsubsection{NFW profile}\label{subsec:analytic_NFW}

The mass function of the NFW distribution 
diverges logarithmically at infinity, requiring the introduction of a cutoff radius $r_c$ to render the total mass finite. Imposing $m(r_c) = M+\MBH$, the mass function can be obtained analytically from the background field 
Eq.~\eqref{TOV_m'}:
\begin{widetext}
\begin{equation}\label{m_NFW}
    m(r) = \left\{\begin{aligned}
    &M+ \MBH + \frac{M}{r+a_0} \left[\frac{(r+a_0)(a_0 + r_c)\log{\left(\frac{r+a_0}{a_0+r_c}\right)} - (a_0 + 2 \MBH) (r-r_c)}{2\MBH - r_c - (a_0 + r_c) \log{\left( \frac{a_0 + 2 \MBH}{a_0+r_c}\right)}} \right] &&\text{for}\quad r<r_c\\
    &M+ \MBH  &&\text{for}\quad r\geq r_c
    \ .
\end{aligned}\right.
\end{equation}
\end{widetext}

The piecewise mass function~\eqref{m_NFW} 
modifies only the source term of the first-order equation for $h_0^{(1)}$, which is otherwise identical to Eq.~\eqref{h0_master_eta_first}, with source term now given by
\begin{widetext}
\begin{equation}\label{source_NFW}
    \mathcal{S}(r) = \left\{ \begin{aligned}
        & - \frac{H_{20}a_0(a_0+r_c)r\left[(2\MBH-r)(8a_0+7r)+8(a_0+r)^2\log\frac{a_0+r}{a_0+2\MBH}\right]}{6(2\MBH-r)(a_0+r)^2\left(r_c-2\MBH+(a_0+r_c)\log\frac{a_0+2\MBH}{a_0+r_c}\right)} &&\text{for}\quad r<r_c \\
        & -\frac{4H_{20}a_0r}{3(r-2\MBH)}&&\text{for}\quad r\geq r_c \ .
    \end{aligned}\right.
\end{equation}
\end{widetext}

Applying the Green's function method as before, the full inhomogeneous solution is given by Eq.~\eqref{h0_Wronskian} with the piece-wise source term Eq.~\eqref{source_NFW}. Since we are interested in the large-radius regime, we focus on the $r>r_c$ region.

Unlike in the Hernquist case, the asymptotic solution obtained for the NFW profile contains no logarithmic terms and can be expressed as a polynomial series\footnote{The integrals in 
Eq.~\eqref{h0_Wronskian} with the source term Eq.~\eqref{source_NFW} contain 
terms proportional to $\text{arctanh}\left(\frac{\MBH}{\MBH-r}\right)$, which admit a polynomial expansion for $r\to\infty$.}. We believe, this is a 
direct consequence of the cutoff, which ensures a strictly 
vacuum exterior and removes the non-compact support responsible for 
the logarithmic contributions in the Hernquist case.

\begin{figure}[H]
\begin{center}
\centering
        \includegraphics[width=1\linewidth]{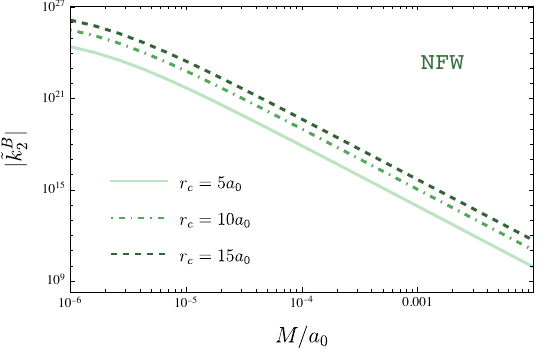}
        \caption{Analytic axial Love numbers $\Tilde{k}_2^B$ for the 
        NFW profile (Eq.~\eqref{k2_full_NFW}) as a function of the compactness $M/a_0$ 
        for different choices of the halo cutoff $r_c$, 
        setting $\MBH=1$ and $a_0=10^6M_{\rm BH}$ and varying $M \in [1,10^4]M_{\rm BH}$. }
        \label{fig:k2_NFW_rc}
\end{center}
\end{figure}

Following the standard matching procedure, we obtain a closed-form expression for the axial TLN of the NFW profile. The full expression, although rather lengthy, is reported in Eq.~\eqref{k2_full_NFW} of 
Appendix~\ref{app:full_TLN}. In the 
limit $a_0,r_c\gg \MBH$, the result simplifies to
\begin{align}\label{k2_analytical_NFW}
    \Tilde{k}^B_2 \sim& -\frac{4a_0^4 M}{\Lscale^5}\nonumber\\
    &+
    \frac{Mr_c^2(3r_c^3-5a_0r_c^2+10a_0^2r_c-30a_0^3)}{15\Lscale^5\left[r_c+(a_0+r_c)\log\frac{a_0}{a_0+r_c}\right]} \ .
\end{align}
As in the Hernquist case, the Love number 
vanishes in the limit $M\rightarrow 0$.
The result is shown in Fig.~\ref{fig:k2_NFW_rc} 
as a function of compactness $M/a_0$\footnote{Note that 
$\tilde{k}_2^B \propto M/(M+M_{\rm BH})^5$ is 
non-monotonic in $M$: it grows linearly for 
$M \ll M_{\rm BH}$ and decreases as $M^{-4}$ for 
$M \gg M_{\rm BH}$. The figures are plotted in the 
range $M \gtrsim M_{\rm BH}$, which lies entirely on 
the decreasing branch, consistent with the $M\to 0$ 
vanishing occurring below the displayed range.},
and three choices of the cutoff radius. 
As $r_c$ increases, the distribution becomes more diluted 
for fixed halo mass $M$, leading to larger Love numbers. 

\subsubsection{Einasto profile}\label{subsec:analytic_Einasto}

For the Einasto profile, the mass function takes 
the analytic form
\begin{align}\label{m_einasto}
    m(r) =& M+ \MBH +\frac{M r^2}{8\MBH^3}\nonumber \\
    &\times\textstyle\left[\frac{2\MBH E_{1-2n}\left(d(\frac{r}{a_0})^{\frac{1}{n}}\right) - r E_{1-3n}\left(d(\frac{r}{a_0})^{\frac{1}{n}}\right)}{E_{1-3n}\left(d\left(\frac{2\MBH}{a_0}\right)^{\frac{1}{n}}\right)-E_{1-2n}\left(d\left(\frac{2\MBH}{a_0}\right)^{\frac{1}{n}}\right)}\right] \ , 
\end{align}
where $E_n(x)$ denotes the exponential integral 
function.

The source term of Eq.~\eqref{h0_master_eta_first} 
for the Einasto profile is considerably more 
involved than in the previous cases. An analytic 
evaluation of the first integral in 
Eq.~\eqref{h0_Wronskian} confirms that no 
logarithmic terms arise from this contribution. 
The second integral, however, does not admit a 
closed-form solution, leaving open the possibility 
that logarithmic terms may still appear in the 
full asymptotic expansion. To address this, we 
evaluate the second integral numerically and 
interpolate the perturbation with a non-linear model 
against both asymptotic forms: the purely polynomial ACMC 
structure~\eqref{ACMC_tphi} and the expansion 
including logarithmic 
corrections~\eqref{h0_expansion_infinity}. The 
Love number is then extracted from the 
coefficients of the interpolation and compared 
with the full numerical integration in the next 
section. 

\begin{figure}[H]
\begin{center}
\centering
        \includegraphics[width=1\linewidth]{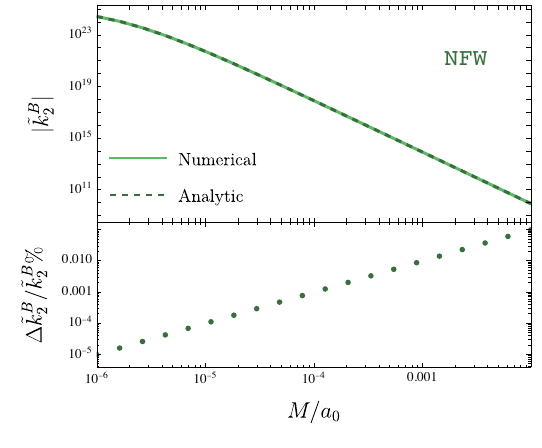}
        \caption{Axial Love numbers $\Tilde{k}_2^B$ for the NFW profile as a function of the 
        compactness $M/a_0$, setting $\MBH=1$, $r_c=5a_0$ and $a_0=10^6M_{\rm BH}$. Top panel: 
        comparison between the numerical solution and the small-compactness approximation 
        (Eq.~\eqref{k2_full_NFW}). Bottom panel: relative difference between the two results.}
        \label{fig:k2_anal_num_NFW}
\end{center}
\end{figure}
%

\subsection{Numerical integration}

In the fully numerical approach, we integrate the 
master Eq.~\eqref{h0_master_Ecluster} from the 
horizon up to a cutoff radius at large distances, beyond 
which the density is set to zero.
We adopt a standard prescription commonly used in the 
study of extended compact configurations, such as boson 
and fermion stars \cite{Delgado:2020udb,Adam_2022,
Cardoso:2017cfl,DeLuca:2022xlz,DelGrosso:2023dmv,
Berti:2024moe,Cannizzaro:2024fpz}. Specifically, 
we define the cutoff radius $R_{99}$ as the 
radius enclosing $99\%$ of the total mass \cite{Berti:2024moe}. 
For $r>R_{99}$, the solution is matched to a 
vacuum exterior out to $R_\text{ext} \gg R_{99}$, 
where the Love numbers $k_\ell^B$ are 
numerically extracted. We have verified that 
varying $R_{99}$ and $R_\text{ext}$ does not 
affect, qualitatively and quantitatively, our 
results. Regularity at the horizon is imposed 
by expanding the perturbation as
\begin{equation}\label{BHhor}
    h_0(r) = \sum_{i=0}^{N} h_0^i (r-2\MBH)^i \ , 
\end{equation}
where we fix $N=5$, which ensures numerical stability. 
The coefficients $h_0^i$ are determined order by order 
by substituting Eq.~\eqref{BHhor} into the master 
equation and expanding around $r=2M_{\rm BH}$. The 
solution is defined up to an overall normalization, 
which is irrelevant for the computation of the Love 
numbers and is therefore fixed to unity.

After the interior integration, we extract 
$h_0(R_{99})$ and $h_0'(R_{99})$, which serve as 
boundary conditions for the exterior vacuum solution. 
In this region, Eq.~\eqref{h0_master_Ecluster} reduces to
\begin{equation}
   \left(1-\frac{2M_\text{tot}}{r}\right) h_0'' 
   - \left(\frac{6}{r^2} - \frac{4M_\text{tot}}{r^3}
   \right) h_0 = 0 \ ,
\end{equation}
with $M_\text{tot} = M+\MBH$. The exterior solution 
depends on two integration constants, which are fixed 
by matching its asymptotic behavior to the ACMC 
expansion~\eqref{ACMC_tphi} at $r=R_\text{ext}$. 
The Love number is then obtained 
as~\cite{Binnington:2009bb,Damour:2009vw,Hinderer:2007mb}
\begin{equation}\label{sigma_formula}
    \Tilde{k}_2^B  
    = \frac{96}{5} \left(\frac{M_\text{tot}}{\Lscale}\right)^5
    \frac{2\xi(y-2) - y + 3}{\mathcal{D}(\xi,y)} \ ,
\end{equation}
where
\begin{align}
    \mathcal{D}(\xi,y) =&\;2\xi\big[2\xi^3(y+1) 
    + 2\xi^2 y + 3\xi(y-1) - 3y + 9\big] \nonumber\\
    &+ 3\big[2\xi(y-2) - y + 3\big]\log(1-2\xi) \ ,
\end{align}
$y=y(R_\text{ext})= R_\text{ext} 
h_0'{}^\text{(ext)}(R_\text{ext})/
h_0^\text{(ext)}(R_\text{ext})$ and 
$\xi= M_\text{tot}/R_\text{ext}$. 

The numerical pipeline is general and can be 
straightforwardly adapted to other spherically 
symmetric backgrounds beyond the dark matter profiles 
considered here. The code used to integrate the field 
equations and compute the perturbations is publicly 
available at \cite{RepoGitHub,DattaGitHub,SGREP_REPO}.

\begin{figure*}
    \centering
    \begin{minipage}{0.49\textwidth}
        \centering
        \includegraphics[width=1\linewidth]{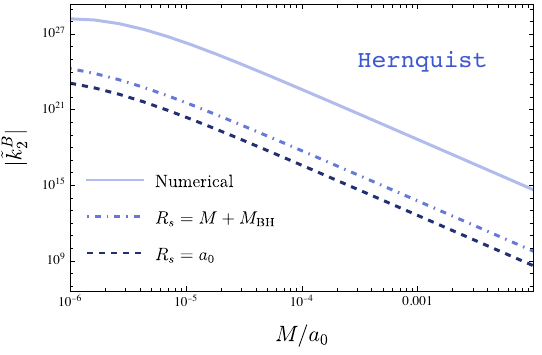}
    \end{minipage}%
    \hfill
    \begin{minipage}[c]{0.49\textwidth}
        \centering
        \includegraphics[width=1\linewidth]{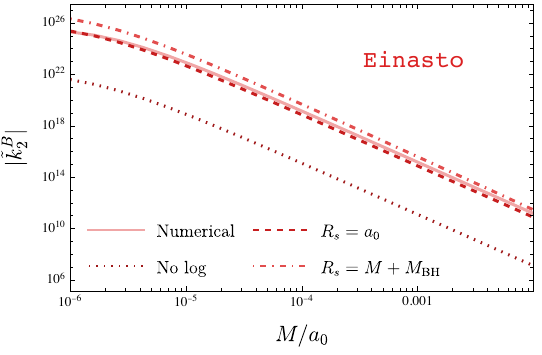}
    \end{minipage}%
    \par
    \centering
    \begin{minipage}[t]{0.49\textwidth}
        \centering
        \caption{Axial Love numbers for the Hernquist profile 
                as a function of the compactness $M/a_0$. The 
                numerical result is compared against the analytic 
                prediction  (Eq.~\eqref{k2_analytical}) obtained by neglecting the logarithmic 
                terms in the asymptotic expansion, without 
                imposing a radial cutoff at large distances 
                from the halo. Different analytic curves 
                correspond to the two standard choices of the 
                arbitrary scale $R_s$ adopted in the literature, 
                namely $R_s=M+\MBH$ and $R_s=a_0$.}
        \label{fig:k2_herquist_analytic_Rs}
    \end{minipage}%
    \hfill
    \begin{minipage}[t]{0.49\textwidth}
        \centering
        \caption{Axial Love numbers for the Einasto profile 
                as a function of the compactness $M/a_0$. The 
                numerical result is compared against predictions 
                obtained by interpolating the perturbation function 
                at large $r$. The interpolation is performed using 
                both a purely polynomial ansatz (labeled ``No log'') 
                and an expansion including logarithmic corrections 
                as in Eq.~\eqref{h0_expansion_infinity}, with the 
                two standard choices $R_s=M+\MBH$ and 
                $R_s=a_0$.}
     \label{fig:k2_Einasto_analytic_Rs}
    \end{minipage}
\end{figure*}

We now compare the analytical small-compactness 
results of Sec.~\ref{Sec:smallCexpansion} with the 
numerical solutions. 
Figures~\ref{fig:k2_anal_num_NFW}--\ref{fig:k2_Einasto_analytic_Rs} 
show this comparison for the NFW, Hernquist, and 
Einasto profiles, fixing $a_0=10^6M_{\rm BH}$, 
varying $M$, and fixing 
$\Lscale = M+\MBH$ \cite{Cardoso:2017cfl}.

All computed Love numbers, both analytical and numerical, are negative; in the plots 
we therefore show their absolute values. Negative axial Love numbers have also been 
reported for neutron stars in the irrotational limit \cite{Damour:2009vw} and for exotic 
compact objects \cite{Chakraborty:2026qru,Cardoso:2017cfl}, 
though the underlying mechanisms differ. The large magnitude reflects the extended and 
weakly bound nature of the matter distribution and is several 
orders of magnitude larger than the Love numbers of 
isolated exotic compact objects with compactness similar 
to those of BHs and neutron 
stars~\cite{Cardoso:2017cfl,Maselli:2017vfi,Chakraborty:2026qru}.

We first discuss the NFW profile. Figure~\ref{fig:k2_anal_num_NFW} shows the configuration with $r_c = 5a_0$. The analytical and numerical results are in excellent agreement, with discrepancies increasing, as expected, at larger compactness. For realistic values ${\cal C}\lesssim 10^{-4}$, the relative difference between the numerical and analytical results is below $10^{-3}\%$. However, even at ${\cal C}\sim 0.1$ the small-compactness approximation reproduces the full solution with a relative accuracy better than $1\%$.

The situation changes significantly for the Hernquist profile. 
Since the analytical Love number~\eqref{k2_analytical} depends 
on the arbitrary scale $R_s$, we compare numerical results 
against two representative choices previously adopted in the 
literature, namely $R_s=M+\MBH$ and $R_s = a_0$. 
Figure~\ref{fig:k2_herquist_analytic_Rs} shows that neither 
choice provides satisfactory agreement with the numerical results.
We interpret this discrepancy as a consequence of the logarithmic terms identified in Eq.~\eqref{h0_expansion_infinity}, which were 
neglected in the matching procedure while deriving Eq.~\eqref{k2_analytical}.
As we show below, imposing a cutoff that restores 
compact support eliminates logarithmic terms, 
restoring a well-defined matching procedure in 
agreement with the numerical results.

A similar situation arises for the Einasto profile, 
with the comparison between numerical and semi analytical 
calculations shown in Fig.~\ref{fig:k2_Einasto_analytic_Rs}. 
While the agreement with the full numerical solution is better 
than for the Hernquist profile, none of the 
interpolations adopted to evaluate the asymptotic 
form of the second integral in Eq.~\eqref{h0_Wronskian} 
reproduces the numerical values satisfactorily. 
This suggests that, as in the Hernquist case, the non-compact 
support of the profile prevents a consistent extraction 
of the Love number without imposing a radial cutoff.

\subsubsection{Small compactness expansion: cutoff prescription}\label{subsec:analytic_Hernquist_cutoff}

The disagreement between the results of the 
small-compactness expansion and the numerical integration for the Hernquist and 
Einasto profiles points to an intrinsic difference between models with a sharp cutoff, beyond which the spacetime is strictly vacuum, and models with 
non-compact asymptotic tails. This distinction is already apparent in the NFW case, where the cutoff is required by the divergence of the total mass and 
naturally removes the logarithmic contributions.

To test whether the cutoff is indeed the relevant ingredient, we return to the semi-analytic calculation for the Hernquist profile of Sec.~\ref{sec:HernAnalytic} and impose a cutoff radius\footnote{
We have called the cutoff radius $r_o$ to 
distinguish it from the cutoff $r_c$ 
used to regularize the divergent mass 
of the NFW profile.} such that 
$\rho(r\ge \rout )=0$. As in the NFW case, 
the mass function becomes piecewise: 
\begin{widetext}
\begin{equation}\label{m_Hernquist_cutoff}
    m(r) = \left\{\begin{aligned}
    &M+\MBH +\frac{M}{(r+a_0)^2}\frac{(a_0+2\MBH)(r-\rout)[2\rout r -2\MBH(r+\rout)+a_0(r+\rout-4\MBH)]}{(\rout-2\MBH)^2}
    &&\text{for}\quad r<\rout\\
    &M+ \MBH  &&\text{for}\quad r\geq \rout \ .
\end{aligned}\right.
\end{equation}
\end{widetext}

Following the same steps as in 
Sec.~\ref{sec:HernAnalytic}, we solve for 
$h_0^{(1)}$ in Eq.~\eqref{h0_master_eta_first} 
with the source term
\begin{align}
    \mathcal{S}(r) = 
        & -\frac{H_{20}}{3}\frac{a_0 (a_0+\rout)^2 r}{(\rout-2\MBH)^2(r+a_0)^3}\nonumber \\
        &\quad\times(4r^2 - 6 \MBH r +5 a_0 r - 8 a_0 \MBH)\ ,
\end{align}
for $r<\rout$, and
\begin{equation}
    \mathcal{S}(r) = -\frac{4H_{20}a_0r}{3(r-2\MBH)}\ ,
\end{equation}
for $r\geq \rout$. The full inhomogeneous solution 
is again given by Eq.~\eqref{h0_Wronskian}. 
Its asymptotic expansion at large $r$ no longer 
contains logarithmic terms and can therefore be 
matched directly to the ACMC 
expansion~\eqref{ACMC_tphi}. The full expression 
for the Love number is given in 
App.~\ref{app:full_TLN}, while in the limit 
$a_0,\rout \gg \MBH$ it reduces to 
\begin{align}\label{k_2_analytical_Hernquist_cutoff}
    \Tilde{k}_2^B \sim&-\frac{4Ma_0 (60a_0^4 + 90a_0^3\rout +20a_0^2\rout^2 - 5a_0\rout^3 +2 \rout^4)}{15\Lscale^5\rout} \nonumber\\
    &+\frac{16M a_0^4(a_0+\rout)^2 \log\left(\frac{a_0+\rout}{a_0}\right)}{15\Lscale^5\rout^2} \ .
\end{align}

\begin{figure*}[htb!]
    \centering
    \begin{minipage}{0.49\textwidth}
        \centering
        \includegraphics[width=1\linewidth]{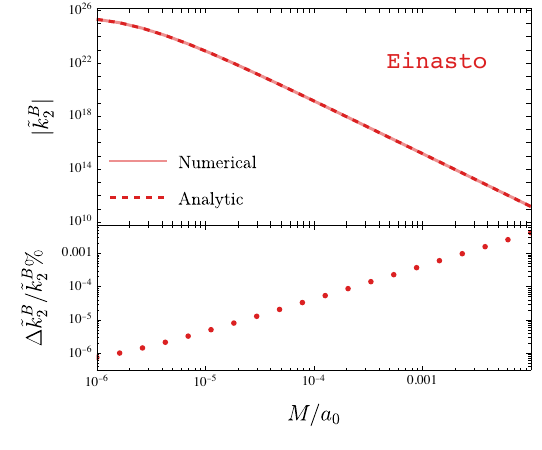}
    \end{minipage}%
    \hfill
    \begin{minipage}[c]{0.49\textwidth}
        \centering
        \includegraphics[width=1\linewidth]{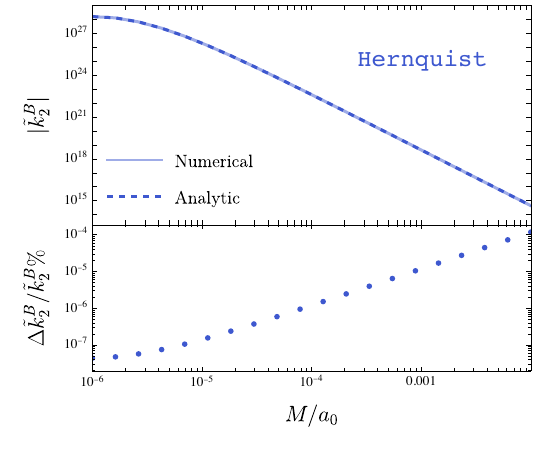}
    \end{minipage}%
    \par
    \centering
    \begin{minipage}[t]{0.49\textwidth}
        \centering
        \caption{Comparison between the numerical 
        solution and the small-compactness expansion 
        for the Hernquist profile with a radial cutoff 
        at $\rout=R_{99}$  (Eq.~\eqref{k2_Hernquist_CutOff_full}). Top panel: axial Love number 
        $\Tilde{k}_2^B$. Bottom panel: fractional 
        difference between the analytic and numerical 
        results. In both panels we fix $\MBH=1$ and 
        $a_0=10^6M_{\rm BH}$, and vary $M \in [1,10^4]M_{\rm BH}$.}
        \label{fig:k2_anal_num_Hernquist}
    \end{minipage}%
    \hfill
    \begin{minipage}[t]{0.49\textwidth}
        \centering
        \caption{Comparison between the numerical 
        solution and the small-compactness expansion 
        for the Einasto profile with a radial cutoff 
        at $\rout=R_{99}$  (Eq.~\eqref{k2_full_Einasto}). Top panel: axial Love number 
        $\Tilde{k}_2^B$. Bottom panel: fractional 
        difference between the analytic and numerical 
        results. In both panels we fix $n=6$, 
        $d=53/3$, $\MBH=1$, and $a_0=10^6M_{\rm BH}$, and vary $M \in [1,10^4]M_{\rm BH}$.}
        \label{fig:k2_anal_num_Einasto}
    \end{minipage}
\end{figure*}

Figure~\ref{fig:k2_anal_num_Hernquist} shows the 
comparison between the small-compactness result 
with the cutoff prescription, assuming $\rout=R_{99}$, 
and the numerical solution. The two approaches agree 
remarkably well, with a fractional difference below 
$10^{-6}\%$ for $M/a_0\lesssim 10^{-4}$. 
Such results remain stable under changes of $\rout$. 
This confirms that the 
logarithmic terms in the Hernquist case are not a 
fundamental feature of the perturbation problem, 
but rather an artifact of the non-compact support 
of the profile. The cutoff prescription restores 
the vacuum boundary conditions required for the 
ACMC matching and yields results fully consistent 
with the numerical integration.

The same analysis can be carried out for the 
Einasto profile. Unlike 
the case without the cut-off, we were able to find 
a closed analytical solution. The full 
analytic expression for the corresponding axial 
Love number is given in Eq.~\eqref{k2_full_Einasto} 
of App.~\ref{app:full_TLN}. As shown in 
Figure~\ref{fig:k2_anal_num_Einasto}, the agreement 
between the analytical and numerical results is 
excellent, consistent with the NFW and cutoff 
Hernquist cases. 

%

\section{Conclusions}

Tidal Love numbers provide a useful characterization 
of the response of a compact object to an external 
gravitational field, and have become an important 
tool for probing the internal structure of 
self-gravitating systems in both gravitational-wave 
physics and strong-field gravity. While much of the 
existing literature has focused on isolated compact 
objects, realistic astrophysical BHs may be 
embedded in nontrivial matter environments. 
Understanding how such surrounding matter modifies 
the tidal response is therefore of both theoretical 
and phenomenological relevance.

In this work we have studied the axial tidal Love 
numbers of a Schwarzschild BH surrounded by 
a spherically symmetric matter distribution modeled 
as an anisotropic fluid. Although, for concreteness, 
we specialized part of the analysis to density 
profiles motivated by dark-matter halos, the 
formalism developed here is more general and applies 
to any spherically symmetric anisotropic medium 
described within the same framework. In this sense, 
our results extend previous studies that had focused 
primarily on specific matter models, and in 
particular on the Hernquist 
profile.

At the formal level, we have clarified certain 
aspects of the structure of axial perturbations 
in anisotropic configurations. 
In particular, 
we have shown that imposing the irrotational 
condition on the fluid four-velocity yields a 
constraint equation for the second velocity 
perturbation entering the problem, whose solutions, 
corresponding to different choices of boundary 
conditions, lead to distinct physical realizations 
of the fluid perturbations.
This observation physically maps to 
different prescriptions that have appeared in the 
literature \cite{Chakraborty:2024gcr}: in the 
Regge--Wheeler formulation, the two families of 
axial Love numbers previously denoted as \textit{``down''} 
and \textit{``up''} correspond, within our framework, 
to the irrotational branch and to the strictly static 
limit, respectively.

We have then studied the problem with two 
complementary approaches. First, we developed a 
small-compactness expansion, which yields 
closed-form analytic expressions for the axial 
Love numbers. Second, we solved the perturbation 
equations numerically, integrating the full system 
without approximation. The comparison between the 
two methods proved especially informative. For the 
NFW profile, the agreement between the analytical 
and numerical results is excellent in the expected 
regime of validity of the expansion. By contrast, 
for the Hernquist and Einasto profiles the analytic 
treatment fails to reproduce the numerical behavior. 
We traced this discrepancy to the appearance of 
logarithmic terms in the asymptotic expansion of 
the perturbation, which originate from the 
non-compact support of the matter distribution. 
To our knowledge, this provides the first clear 
physical interpretation of logarithmic contributions 
that had been noticed previously but not fully 
understood.

Motivated by this observation, we introduced a 
cutoff prescription that enforces a strictly vacuum 
exterior region also in the semi-analytic treatment. 
Once this is done, the logarithmic terms disappear 
and the agreement between the small-compactness 
expansion and the numerical solution is restored. 
More broadly, this result 
highlights the importance of carefully controlling 
the asymptotic structure of the background when 
defining tidal observables in black-hole spacetimes 
dressed by matter.

The most immediate extension is the study of polar TLNs, which are expected
to provide the dominant contribution to gravitational-wave signals from
coalescing binaries and whose perturbation equations have a richer
structure~\cite{Chakraborty:2024gcr}. 
It would also be interesting to assess the sensitivity of the results to
the choice of matter model, by extending the analysis to other spherically
symmetric distributions beyond those considered here, and to investigate
the prospects of future gravitational-wave detectors to constrain or
distinguish between different matter environments through measurements 
of tidal parameters. 
In particular, it would be interesting to study the 
dynamical evolution of the tidal deformability in the 
presence of matter, accounting for tidal interactions 
and mass accretion that could deplete the environment 
during the inspiral~\cite{DeLuca:2022xlz}.

\section*{Acknowledgments}
S. D'Onofrio thanks Claudio Gambino for useful discussions. 
This work is funded by MCIN/AEI/10.13039/501100011033 and 
FSE+, reference PRE2021-098098 (S. D'Onofrio).
S.D. acknowledges financial support from MUR, PNRR - Missione 4 - Componente 2 - 
Investimento 1.2 - finanziato dall'Unione europea - NextGenerationEU 
(cod. id.: SOE2024\_0000167, CUP:D13C25000660001). 
A.M.~acknowledges financial support from MUR PRIN 
Grants No.~2022-Z9X4XS and No.~2020KB33TP.
We thank Sumanta Chakraborty and Chiranjeeb Singha for useful discussions.

\appendix

\section{Full expressions for the axial Love numbers}\label{app:full_TLN}

In this appendix, we provide the full analytic 
expressions for the axial Love numbers associated with the 
profiles considered in the main text, when a 
radial cutoff to the density is applied. 
For the NFW profile, we obtain
\begin{widetext}
    \begin{align}\label{k2_full_NFW}
    \Tilde{k}_2^B{}_\text{NFW} = -\Bigg[&12 a_0^3 M (5 a_0+8 \MBH) (a_0+r_c) \log \left(\frac{a_0+2 \MBH}{a_0+r_c}\right)\nonumber\\ 
    &-M (2 \MBH-r_c) \left(4 a_0 (a_0+\MBH) \left(15 a_0^2-6 a_0 \MBH+2 \MBH^2\right)+2 r_c \left(15 a_0^3+14 a_0^2 \MBH-6 a_0 \MBH^2+4 \MBH^3\right) \right.\nonumber\\
    &\left.+r_c^3 (5 a_0+2 \MBH)-2 r_c^2 (5 a_0-2 \MBH) (a_0+\MBH)-3 r_c^4\right)\Bigg] \nonumber \\
    \times &
    \left(15 \Lscale^5 \left((a_0+r_c) \log \left(\frac{a_0+2 \MBH}{a_0+r_c}\right)-2 \MBH+r_c\right)\right)^{-1} \ .
\end{align}
\end{widetext}
While the analytical expression for the Hernquist profile without 
cut-off has been given in the main text (Eq.~\eqref{k2_analytical}), 
the full expression with a radial cut-off $\rout$ has the lengthy expression 
\begin{widetext}
\begin{align}\label{k2_Hernquist_CutOff_full}
    \Tilde{k}_2^B{}_\text{Hernquist} = &\frac{4 M}{15 \Lscale^5 (\rout-2 \MBH)^2} \Bigg\{12 a_0^2 (a_0+2 \MBH) (5 a_0+6 \MBH) (a_0+\rout)^2 \log \left(\frac{a_0+\rout}{a_0+2 \MBH}\right)\nonumber\\
    &+(2 \MBH-\rout) \bigg[60 a_0^5+6 a_0^4 (22 \MBH+15 \rout)+4 a_0^3 \left(8 \MBH^2+52 \MBH \rout+5 \rout^2\right)\nonumber\\
    &+a_0^2 \left(-8 \MBH^3+60 \MBH^2 \rout+54 \MBH \rout^2-5 \rout^3\right)-2 a_0 \rout (2 \MBH-\rout)^3\nonumber\\
    &-4 \MBH \rout^2 (2 \MBH-\rout) (\MBH+\rout)\bigg]\Bigg\} \ .
\end{align}
\end{widetext}
The complexity of the Einasto mass function prevents a closed-form 
evaluation of the second integral in Eq.~\eqref{h0_Wronskian} without 
a radial cutoff. With a cutoff $\rout$, however, the source term 
reduces to the vacuum expression for $r\geq \rout$, and the full 
analytical solution can be obtained. The resulting axial TLN is
\begin{widetext}
\begin{align}\label{k2_full_Einasto}
\Tilde{k}_2^B{}_\text{Einasto}& = \frac{M}{90 \sqrt[6]{a_0} \Lscale^5 }\nonumber\\
\times\Bigg\{&\textstyle6784 \sqrt[6]{2} \MBH^{43/6} \left(E_{-42}\left(\frac{53}{3} \sqrt[6]{\frac{2\MBH}{a_0}}\right)-E_{-36}\left(\frac{53}{3} \sqrt[6]{\frac{2\MBH}{a_0}}\right)-E_{-18}\left(\frac{53}{3} \sqrt[6]{\frac{2\MBH}{a_0}}\right)+E_{-12}\left(\frac{53}{3} \sqrt[6]{\frac{2\MBH}{a_0}}\right)\right)\nonumber\\
&\textstyle-2304 \sqrt[6]{a_0} \MBH^7 \left(11 E_{-41}\left(\frac{53}{3} \sqrt[6]{\frac{2\MBH}{a_0}}\right)-10 E_{-35}\left(\frac{53}{3} \sqrt[6]{\frac{2\MBH}{a_0}}\right)-3 E_{-17}\left(\frac{53}{3} \sqrt[6]{\frac{2\MBH}{a_0}}\right)+2 E_{-11}\left(\frac{53}{3} \sqrt[6]{\frac{2\MBH}{a_0}}\right)\right)\nonumber\\
&\textstyle+106 \MBH \rout^{37/6} \left(E_{-36}\left(\frac{53}{3}\sqrt[6]{\frac{\rout}{a_0}}\right)-E_{-12}\left(\frac{53}{3}\sqrt[6]{\frac{\rout}{a_0}}\right)\right)+72 \sqrt[6]{a_0} \MBH \rout^6 \left(E_{-11}\left(\frac{53}{3}\sqrt[6]{\frac{\rout}{a_0}}\right)-5 E_{-35}\left(\frac{53}{3}\sqrt[6]{\frac{\rout}{a_0}}\right)\right)\nonumber\\
&\textstyle+53 \rout^{43/6} \left(E_{-18}\left(\frac{53}{3}\sqrt[6]{\frac{\rout}{a_0}}\right)-E_{-42}\left(\frac{53}{3}\sqrt[6]{\frac{\rout}{a_0}}\right)\right)+18 \sqrt[6]{a_0} \rout^7 \left(11 E_{-41}\left(\frac{53}{3}\sqrt[6]{\frac{\rout}{a_0}}\right)-3 E_{-17}\left(\frac{53}{3}\sqrt[6]{\frac{\rout}{a_0}}\right)\right)\Bigg\}\nonumber\\
\times\Bigg\{ & \textstyle\left(8 \MBH^3 \left(E_{-17}\left(\frac{53}{3} \sqrt[6]{\frac{2\MBH}{a_0}}\right)-E_{-11}\left(\frac{53}{3} \sqrt[6]{\frac{2\MBH}{a_0}}\right)\right)+2 \MBH \rout^2 E_{-11}\left(\frac{53}{3}\sqrt[6]{\frac{\rout}{a_0}}\right)-\rout^3 E_{-17}\left(\frac{53}{3}\sqrt[6]{\frac{\rout}{a_0}}\right)\right)\Bigg\}^{-1} \ ,
\end{align}
\end{widetext}
where $E_n(x)$ denotes the exponential integral function.

\addtocontents{toc}{\setcounter{tocdepth}{-10}} 
\newpage
\phantomsection

\bibliographystyle{apsrev4-2}

\bibliography{Bibliography}

\end{document}